\documentclass[aps,prd,twocolumn,showkeys,nofootinbib,amsmath,amssymb,superscriptaddress]{revtex4-2}
\usepackage{tikz}
\usetikzlibrary{arrows.meta,
                positioning,
                quotes
                }
\usepackage{dsfont}
\usetikzlibrary{positioning}
\usepackage{mathtools} 
\newcommand{\tikznode}[2]{\relax
    \ifmmode%
    \tikz[remember picture,baseline=(#1.base),inner sep=0pt] \node (#1) {$#2$};
    \else
    \tikz[remember picture,baseline=(#1.base),inner sep=0pt] \node (#1) {#2};%
    \fi}

\usepackage{xcolor}
\definecolor{nicegreen}{rgb}{0.07, 0.564, 0.04}
\usepackage{framed}
\usepackage{comment}
\usepackage{xspace}
\usepackage{graphicx}
\usepackage{dcolumn}
\usepackage{bm}
\usepackage{tikz-cd}
\usepackage[normalem]{ulem}
\usepackage{slashed}

\usepackage[colorlinks]{hyperref}
\hypersetup{
  breaklinks=true,
  colorlinks=true,
  linkcolor=blue,
  citecolor=red,
  urlcolor=magenta
}

\def\LevelSep{25mm}   
\def\HalfSep{55mm} 

\newcommand{\diff}{d}

\DeclarePairedDelimiterX\Braket[2]{\langle}{\rangle}{#1 \delimsize\vert #2}

\begin{document}

\title{$\kappa$-entropic statistical paradigm for relativistic corrections to the Heisenberg principle}

\author{G.~G.~Luciano}
\email{giuseppegaetano.luciano@udl.cat}
\affiliation{Department of Chemistry, Physics and Environmental and Soil Sciences, Polytechnic School, University of Lleida, Av. Jaume II, 69, 25001 Lleida, Spain
}

\author{J.~Gin\' e}
\email{jaume.gine@udl.cat}
\affiliation{Department of Mathematics, Polytechnic School, University of Lleida, Av. Jaume II, 69, 25001 Lleida, Spain}

\author{D.~Chemisana}
\email{daniel.chemisana@udl.cat}
\affiliation{Department of Chemistry, Physics and Environmental and Soil Sciences, Polytechnic School, University of Lleida, Av. Jaume II, 69, 25001 Lleida, Spain
}

\date{\today}

\begin{abstract}
The Heisenberg position–momentum uncertainty relation is a cornerstone of quantum mechanics. However, its standard formulation is not fully consistent with special relativity. While partial understanding has been achieved in the ultra-relativistic regime, a comprehensive description is still lacking, particularly in the intermediate velocity domain, where particle speeds remain well below the speed of light yet relativistic corrections are expected to become appreciable. This regime constitutes the most promising arena for experimentally probing relativistic modifications of quantum uncertainty.
By adopting a variational approach, in this work we derive a relativistic extension of the Heisenberg algebra within the framework of $\kappa$-deformed Kaniadakis statistics. The latter emerges from the application of the Maximum Entropy Principle to Kaniadakis entropy, a one-parameter generalization of the Boltzmann–Gibbs–Shannon entropy naturally induced by Lorentz transformations. We investigate the physical implications of the resulting uncertainty relation, deriving constraints on the Kaniadakis parameter from precision measurements of the fine-structure constant and confronting our construction with other extensions discussed in the recent literature.
\end{abstract}

\maketitle

\section{Introduction}

Reconciling Quantum Mechanics (QM) and General Relativity (GR) remains one of the central open problems in theoretical physics. Despite numerous candidate approaches~\cite{Carlip,Hosse,AC,RovelliRev,Addazi}, a complete and experimentally verified theory of quantum gravity (QG) is still lacking.

Among the most robust and widely shared expectations of QG theories is the emergence of a minimal length, typically associated with the Planck scale. In phenomenological models, this feature is commonly incorporated through a deformation of the Heisenberg Uncertainty Principle (HUP) into a Generalized Uncertainty Principle (GUP)~\cite{Amati,Gross,Amati2,Scardigli}, which accommodates a nonvanishing minimal position uncertainty~\cite{KMM,DasPrl,Jizba,Isi,FrPan,BossoCohe,ScardLamb,ScardLuc,Wagner2021,BH1,BH3,BH4,BH5,BH5bis,BH6,BH6bis,BH10,PetrozN,Iorio,Bruk,GravBar,ScardCas,Gauge,Husain,BossoLuciano,Casadio:2022opg,Singh,ChenGUP}. Similar generalizations also arise in extended formulations
of QM, such as non-associative quantum theories~\cite{Bojo}, as well as in
models defined on noncommutative spacetimes~\cite{Dimit}.

While identifying the scale at which QG effects become significant, the introduction of a universal minimal length appears at odds with the principles of GR, where distances are defined through a dynamical metric rather than a fixed geometric background (see, e.g., Snyder's noncommutative spacetime model for an early attempt at reconciling relativistic invariance with a fundamental length scale~\cite{Snyder,Minwalla}). This tension has motivated several attempts to develop a generally covariant formulation of the HUP. However, many such constructions either leave unresolved conceptual issues  or even introduce additional inconsistencies~\cite{BossoRev}.
 
\begin{figure*}[t]
  \centering
  \resizebox{1\textwidth}{!}{
  \begin{tikzpicture}[
    auto,
    > = Stealth,
    box/.style = {
      draw=gray, very thick,
      minimum height=15mm,
      minimum width=57mm,
      text width=74mm,
      font=\normalsize,
      align=center
    },
    every edge/.style = {draw, ->, very thick},
    every edge quotes/.style = {font=\small, align=center, inner sep=1pt}
  ]

  \def\LevelSep{20mm}   
  \def\HalfSep{50mm}

  \node (HUPtop) [box]
  {\textbf{Non-relativistic regime (HUP)} \\ \vspace{2mm}
  $[x,p]=i\hbar$ \, (CS $\rightarrow$ Gaussian WP)};

  \node (MID) [coordinate, below=\LevelSep of HUPtop] {};

  \node (RUPmid) [box, xshift=-\HalfSep] at (MID)
{\textbf{\small Non-relativistic gravitational regime (GUP)} \\ \vspace{2mm}
  \small $[x,p]=i\hbar\!\left(1+\beta \frac{p^2}{m_p^2c^2}\right)$ \, (CS $\rightarrow$ $q$-Gaussian WP)};

  \node (GUPmid) [box, xshift=\HalfSep] at (MID)
  {\textbf{Special-relativistic regime (RUP)} \\ \vspace{2mm}
  Eq.~\eqref{RUP} \, (CS $\rightarrow$ $\kappa$-Gaussian WP)};

  \node (GRbot) [box, below=\LevelSep of MID]
  {\textbf{General-relativistic regime} \\ \vspace{2mm}
  \large{\textbf{?}}};

  \draw (HUPtop.south west) edge [
    line width=2pt,
    "{\bfseries Gravity}"'{yshift=0mm}
  ] (RUPmid.north);

  \draw (HUPtop.south east) edge [
    line width=2pt,
    "{\bfseries Special relativity\\{\normalfont\small (our model)}}"{yshift=-4mm}
  ] (GUPmid.north);

  \draw (RUPmid.south) edge [
    line width=2pt
  ] (GRbot.north west);

  \draw (GUPmid.south) edge [
    line width=2pt
  ] (GRbot.north east);

  \draw[dashed, line width=1.2pt, ->]
    (HUPtop.south) -- (GRbot.north);

  \end{tikzpicture}
  }
  \caption{Diagrammatic map of uncertainty relations and their connections to deformed statistics. Solid arrows denote extensions of the HUP that can be addressed within potentially controlled frameworks, while the dashed arrow indicates the more challenging direct transition to the GR regime.}
  \label{figure}
\end{figure*}
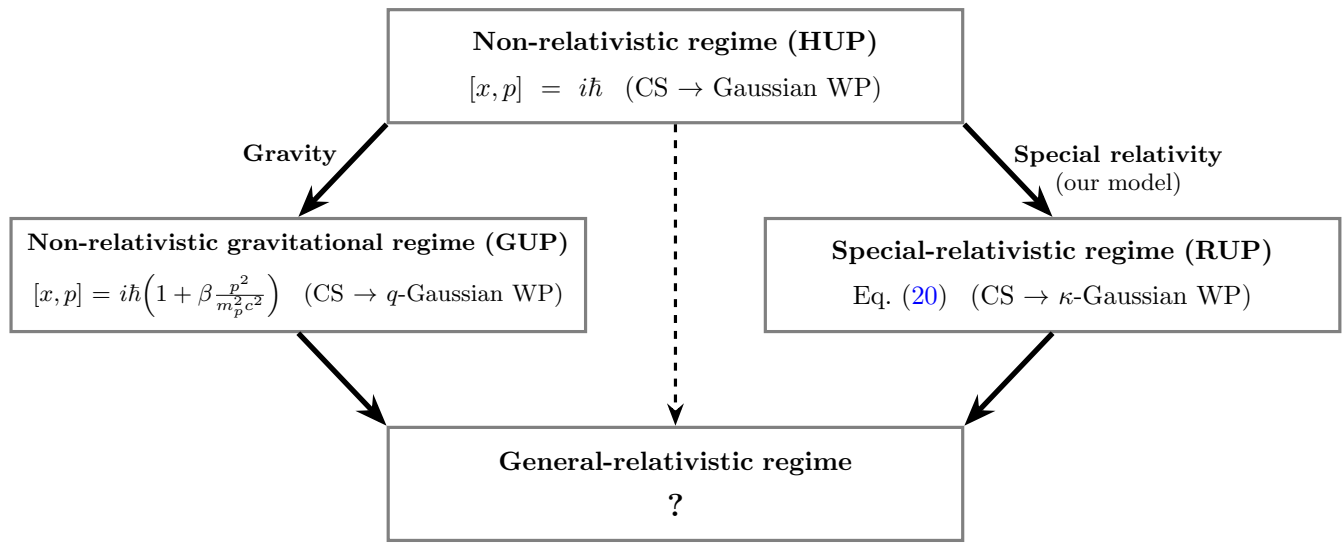

In view of this impasse, a more conservative strategy appears warranted. Before addressing the interplay between quantum and gravity, it is natural to first clarify how the HUP is modified when \emph{special-relativistic} (SR) effects are taken into account (see Fig.~\ref{figure}).

The interface between SR and QM is described by special-relativistic quantum field theory. Nevertheless, the fate of the HUP remains quite obscure in the transition from the Galilean formulation of QM to its relativistic counterpart (see also~\cite{Mamon} for a recent discussion on the incompatibilities between SR and QM). 

In particular, since \emph{i}) the standard HUP relates uncertainties in the simultaneous measurements of position and momentum, and \emph{ii}) position ceases to be a well-defined observable in special-relativistic quantum field theory, it is commonly believed that the traditional position–momentum uncertainty relation abruptly loses its operational meaning as the SR regime is approached.

Supported by the analyses of~\cite{Land,Newt,Freidel,BossoTodo,GAC}, it is instead natural to expect the existence of an intermediate regime in which SR effects are appreciable enough to require careful consideration, yet not so dominant as to invalidate the conceptual structure of the HUP. In this regime, the most plausible scenario is that the HUP paradigm remains valid, while incorporating SR effects through the appearance of additional corrections. Moreover, it is precisely this domain that is best suited for experimentally probing the interplay between QM and SR.

To the best of our knowledge, the first attempt to study a relativistic formulation of the HUP was presented in~\cite{Land}. By exploiting the idea that any position measurement cannot generate energy fluctuations capable of creating new quanta, it was argued that the 
localization precision of a particle at rest is bounded from below by its Compton wavelength. Subsequent investigations have explored deformations of the Heisenberg description based either on the definition of observables and the Feynman propagator in the single-particle SR Hilbert space~\cite{Newt,Freidel}, or on the properties of the probability distribution associated with the wave function~\cite{GAC}. 

All these constructions are phenomenologically implemented at the level of the uncertainty relation. In this work, we instead develop a more general formalism that enables a relativistic extension of the HUP directly at the algebraic level. Building on recent results~\cite{JizbaLuc,Jizba:2023ygi}, which establish a correspondence between the coherent states (CSs) of the GUP and the probability amplitude of the Tsallis distribution for free nonrelativistic particles~\cite{Tsallis}, we naturally extend this framework by seeking a deformation of the canonical commutator whose CSs are compatible with the relativistic generalization of the Maxwell–Boltzmann (MB) distribution. The latter is provided by the $\kappa$--deformed distribution that maximizes the Kaniadakis entropy~\cite{Kania0,Kania1,Kania2,LucianoRev}, a one-parameter generalization of the Boltzmann-Gibbs-Shannon entropy, recovered in the limit $\kappa \rightarrow 0$.

This construction naturally leads to a self-consistent special-relativistic formulation of the HUP in the quasi-classical limit, in close analogy with the way the non-relativistic GUP emerges from Tsallis thermostatistics~\cite{JizbaLuc}. Henceforth, we shall refer to this model as the Relativistic Uncertainty Principle (RUP).  In the following, we investigate the resulting corrections to the HUP and compare our results with other theoretical and phenomenological derivations of relativistic uncertainty relations proposed in the recent literature.
Since the present construction modifies the canonical commutation relation within a single-particle QM framework, rather than introducing a relativistic position operator in the sense of Refs.~\cite{Newt,Freidel}, the well-known causality issues associated with relativistic localization do not directly arise in the present setting.

The remainder of the paper is organized as follows. In the next section, we fix the notation by reviewing the formalism of Kaniadakis statistics. Section~\ref{RHUP} is devoted to the derivation of the RUP, while in Section~\ref{App} we discuss some physical implications of our results and compare our construction with other relativistic extensions of the uncertainty principle.
Finally, conclusions and outlook are presented in Section~\ref{D&C}. 
The paper is complemented by two appendices containing additional technical details of the analysis. 
Unless otherwise specified, we set the Boltzmann constant $k_B = 1$, while keeping Planck’s constant $\hbar$ and the speed of light $c$ explicit.

\section{Kaniadakis Statistics}
\label{KS}

Let us outline the mathematical and physical foundations of Kaniadakis statistics, which provide the theoretical framework underlying our construction. For a comprehensive treatment, we refer the reader to~\cite{Kania0,Kania1,Kania2}.
  
It is well known that the Maxwell-Boltzmann (MB) distribution is commonly taken as a foundational element of classical statistical mechanics, rather than being derived from it. More precisely, it can be regarded as emerging from Newtonian mechanics, as supported by numerical simulations of classical molecular dynamics. A natural question then arises as to whether the MB distribution is also recovered within a framework in which the microscopic dynamics are governed by the laws of SR.

This issue has been thoroughly investigated in~\cite{Kania0,Kania1,Kania2}, motivated by the observation that cosmic rays - arguably the most striking example of relativistic particles - exhibit power-law asymptotic spectra, in contrast with the exponential behavior predicted by the MB distribution. Similar features have been observed in other relativistic systems, such as plasmas in superthermal radiation fields~\cite{plasma} and nuclear collisions~\cite{NC}, pointing to the necessity of adopting an entropy measure that is manifestly different from the standard Boltzmann-Gibbs-Shannon one (see~\cite{LucianoRev,Rev2} for further applications of Kaniadakis entropy in cosmology and astrophysics).

In Refs.~\cite{Kania0,Kania1,Kania2}, it was shown that consistency with SR composition laws (and, in particular, Lorentz-transformation properties) leads to the following one-parameter deformed entropy:
\begin{equation}
    \label{KE}
S_{\kappa} = -\sum_{i} n_i \ln_{\kappa} n_i \,,\,\, \quad\,\, \ln_{\kappa}(y) \equiv \frac{y^{\kappa} - y^{-\kappa}}{2\kappa} \,.
\end{equation}
The generalized Boltzmann factor is obtained by maximizing the entropy \eqref{KE} under the standard constraints of normalization and fixed mean energy. This procedure yields the stationary distribution for the $i$-th energy level in the form $n_i \propto \exp_{\kappa}(-\beta E_i)$ (we set the chemical potential $\mu=0$), where
\begin{equation}
\label{expk}
\exp_{\kappa}(y) \equiv \left(\sqrt{1 + \kappa^{2} y^{2}} + \kappa y \right)^{1/\kappa},
\end{equation}
and $1/\beta = \sqrt{1 - \kappa^{2}}\,T$ denotes the modified inverse temperature of the system. The dimensionless parameter $-1<\kappa<1$ quantifies deviations from ordinary MB statistics, which is recovered in the classical limit $\kappa \to 0$. Since $\exp_{\kappa}(y)$ is even under $\kappa \rightarrow -\kappa$, we restrict the analysis to $\kappa > 0$.

To connect the above formalism with physical observables, let us introduce the following auxiliary dimensionless functions \cite{Kania2}:
\begin{eqnarray}
\label{ut}
u(\tilde p) &=& \frac{\tilde p}{\sqrt{1+\kappa^2 \tilde p^2}}, \\[2mm]
\mathcal{W}(\tilde p) &=& \frac{1}{\kappa^2}\left(\sqrt{1+\kappa^2 \tilde p^2}-1\right), \\[2mm]
\varepsilon(\tilde p) &=& \frac{1}{\kappa^2}\sqrt{1+\kappa^2 \tilde p^2},
\end{eqnarray}
where $\tilde p$ denotes the momentum variable, and $u$, $\mathcal{W}$, and $\varepsilon$ correspond to the velocity, kinetic energy, and total energy of the system, respectively.

The physical velocity $v$, momentum $p$, and total energy $E$ can be defined through the scaling relations \cite{Kania2}
\begin{equation}
\frac{v}{u}
=
\frac{p}{m \tilde p}
=
\sqrt{\frac{E}{m \varepsilon}}
=
\kappa c
\equiv v_* \,,
\label{auxquan}
\end{equation}
where $m$ denotes the rest mass and $c$ the speed of light. The corresponding physical kinetic energy is given by $W = E - mc^2$.

Using the definitions~\eqref{ut}--\eqref{auxquan}, one can verify that the standard special-relativistic expressions for momentum and energy are recovered \cite{Kania2}. Moreover, in order for the above relations to retain physical meaning in the Galilean regime, it must be imposed that the characteristic velocity $v_*$ remain finite in the joint limit $c \to \infty,\,\kappa \to 0$.

Before turning to the formulation of a HUP consistent with the Kaniadakis framework, we recall that other relativistic generalizations of the MB distribution have been proposed. Among these, the Maxwell-J\"uttner velocity distribution~\cite{MJ} represents the earliest attempt at constructing a relativistic statistical theory. Nevertheless, this model is obtained by naively substituting the relativistic energy–velocity relation into the classical MB factor, yielding a distribution that still maximizes the standard Boltzmann-Gibbs-Shannon entropy. By contrast, as discussed above, the Kaniadakis distribution~\eqref{expk} follows from an \emph{ab initio}, fully relativistic entropy functional~\eqref{KE}. 

\section{Relativistic Uncertainty Principle}
\label{RHUP}

Having established the relativistic statistical framework, we now examine its QM implications. In particular, we investigate how the $\kappa$-deformed structure translates into a modification of the Heisenberg algebra, leading to a relativistic generalization of the HUP.
This perspective parallels the GUP, where deformed commutation relations imply a minimal length scale that also emerges from entropy corrections, notably through the black hole entropy–area relationship \cite{Kim}.

\subsection{HUP and Boltzmann–Gibbs Statistics}
Let us briefly recall the statistical interpretation of the HUP. Consider the canonical commutator $[x,p]=i\hbar$, which implies the Robertson inequality 
$\Delta x\,\Delta p \geq \hbar/2$. Here, $x$ and $p$ denote the usual position and momentum operators, 
while $\Delta x$ and $\Delta p$ represent their corresponding uncertainties, defined as the standard deviations. 

By considering, for instance, the momentum representation, it is well known that the minimum-uncertainty condition leads to a first-order differential equation whose normalized solution is given by the Gaussian wave packet
$\psi(p) \propto \exp\!\left(-\frac{p^2}{2\,\Delta p^2}\right)$, up to an irrelevant global phase. Here we have restricted our attention to symmetric states with vanishing expectation values of position and momentum \cite{cohen}.  These states coincide with the well-known Glauber CSs and represent minimum-uncertainty solutions of the HUP.

From a statistical perspective, the Gaussian distribution arises from the maximization of the Boltzmann–Gibbs–Shannon entropy $S=-\int \rho(p)\,\ln \rho(p)\, dp,$, under the constraints of normalization and fixed variance. Indeed, this procedure yields $\rho(p) \propto \exp\!\left(-\sigma p^{2}\right)$, where $\sigma$ is a Lagrange multiplier determined by the imposed constraints. In this sense, the coherent states of the HUP naturally emerge within the Boltzmann-Gibbs statistical framework.

\subsection{GUP and Tsallis statistics}
\label{TGUP}
In recent decades, numerous studies in QG have converged on the idea that the HUP should be modified at the Planck scale to accommodate the emergence of a minimal length. The resulting Generalized Uncertainty Principle (GUP) has been widely applied across a broad range of contexts, including black hole physics, cosmology and quantum theory~\cite{KMM,DasPrl,Jizba,FrPan,BossoCohe,ScardLamb,ScardLuc,Wagner2021,BH1,BH3,BH4,BH5,BH5bis,BH6,BH6bis,BH10,PetrozN,Iorio,Bruk,GravBar,ScardCas,Gauge,Husain,BossoLuciano,Casadio:2022opg,Singh,GUPMA,GUPCasScard}.

In this context, recent studies at the interface between QM and statistical physics~\cite{JizbaLuc,Jizba:2023ygi,Shababi,LucGUP} have uncovered a connection between the GUP - at least in its commonly adopted quadratic form $\Delta x \Delta p \ge \frac{\hbar}{2}\left(1+\beta\, \Delta p^{2}/m_p^2\right)$~\cite{KMM} - and non-extensive Tsallis statistics~\cite{Tsallis}. The latter generalizes the MB framework to systems characterized by long-range interactions and correlations, including gravitational ones~\cite{Tsallis2,PlastinoRocca}.

In particular, it has been shown that the CSs associated with the GUP coincide in momentum space with the probability amplitudes of the $q$-Tsallis distribution for free non-relativistic particles, namely $\psi_q(p)\propto \left[1 - (1-q)\,\frac{p^{2}}{2m_p} \right]^{\frac{1}{2(1-q)}}$~\cite{Tsallis}, provided that the non-extensivity parameter $q$ is appropriately mapped onto the GUP deformation parameter $\beta$ (i.e., $\beta\propto q-1$)~\cite{JizbaLuc,Jizba:2023ygi}.

This correspondence constitutes a non-extensive generalization of the BG framework. Indeed, the $q$-Gaussian states arise from the maximization of the Tsallis entropy $S_q = \frac{1}{1 - q} \left( \int \rho^q(p)\, dp - 1 \right)$ under constraints analogous to normalization and fixed variance. In the limit $q \to 1$, Tsallis statistics reduces to BG statistics, the $q$-Gaussian converges to a standard Gaussian, and the GUP reduces to the HUP, yielding the corresponding minimum-uncertainty Gaussian wave packets.

Taken together, these results indicate that the combined use of GUP CSs and Tsallis entropy provides a natural statistical framework for exploring the semiclassical regime of GUP-deformed QM. Conceptually, this approach follows a top-down logic, whereby the deformation of the uncertainty principle is introduced at the quantum-gravitational level and its statistical implications are subsequently derived.

\subsection{RUP and Kaniadakis statistics}
Motivated by these considerations, we reverse the above logic and reconstruct relativistic corrections to the HUP starting from the $\kappa$-deformed statistical framework. Specifically, we adopt a bottom-up approach in which the minimum-uncertainty states are required to coincide with the $\kappa$-distribution~\eqref{expk}, thereby fixing the corresponding deformation of the Heisenberg algebra. The generalized uncertainty relation then follows from the resulting commutation structure. 

This construction provides an effective algebraic realization of a relativistically deformed uncertainty relation. Although guided by the coherent-state formalism, the resulting framework defines a consistent extension of the Heisenberg structure, whose physical viability will be corroborated a posteriori through comparison with existing phenomenological approaches.

At the operational level, our formalism is implemented through a deformation of the canonical commutation relation of the form
\begin{equation}
\label{RelatUP}
[x,p] = i\hbar\, f(p)\,,
\end{equation}
where $f(p)$ is a positive real function to be determined.

In turn, the Robertson inequality gives the
generalized uncertainty relation~\cite{Robertson}
\begin{equation}
\label{RelatHUP2}
 \Delta x\, \Delta p\ge \left|\frac{1}{2i}{\langle[x,p]\rangle}\right|
    = \frac{\hbar}{2} \langle f(p)\rangle\,,
\end{equation}
for any state in the domain of the operators involved. Standard QM is recovered when $f(p)\to1$, which is expected in the nonrelativistic regime. 

Notably, the assumption that the deformation \eqref{RelatUP} depends only on momentum is supported by previous analyses. In particular, in~\cite{GAC} corrections to the uncertainty relation originate from the relativistic dispersion relation, which modifies the probability distribution of the wave function in momentum space. Accordingly, deviations from the standard HUP reflect the relativistic kinematics encoded in the energy–momentum relation, and are therefore naturally expressed as momentum-dependent corrections. Likewise, the relativistic uncertainty relation proposed in~\cite{Putra} is formulated in terms of the average group velocity of the wave packet, which depends on momentum through the relativistic dispersion relation.

For computational convenience, let us introduce the auxiliary operator $q$ such that
\begin{equation}
\label{canpq}
    [q, p]=i\hbar\,.
\end{equation}
This operator should not be confused with the non-extensivity parameter $q$ appearing in Tsallis statistics, as discussed in Sec.~\ref{TGUP}. 

Assuming that $x \equiv x(q,p)$ is a differentiable operator function, its dependence on the auxiliary variable $q$ can be determined from the commutation relation with $p$. In particular, using the canonical identity
\begin{equation}
    [x(q,p),\, p] = [q,\, p] \, \frac{\partial x(q,p)}{\partial q}=i\hbar \, \frac{\partial x(q,p)}{\partial q}\,.
\end{equation}
it follows that
\begin{equation}
   \frac{\partial x(q,p)}{\partial q} = f(p)\,.
\end{equation}
In order to make the action of the operator \(x\) explicit in terms of \(q\), 
one must adopt a specific operator-ordering prescription. For instance, 
one may choose one of the following representations:
\begin{equation}
\label{threerep}
x_1 = f(p) q, \,\quad\,
x_2 = q f(p), \,\quad\,
x_3 = \frac{1}{2} \left[f(p) q + q f(p)\right],
\end{equation}
where we have imposed the condition $x(q=0,p)=0$. Clearly, the final physical 
results must not depend on this choice~\cite{BossoLuciano}.

In Ref.~\cite{Land}, a relativistic extension of the HUP is formulated by requiring that physically meaningful position measurements do not trigger particle creation. This leads to an effective lower bound on the position uncertainty of a particle at rest, of the order of its Compton wavelength. In close analogy with the GUP framework~\cite{KMM}, the existence of such a minimal uncertainty implies that the Heisenberg algebra cannot be represented in terms of position eigenstates, and therefore does not admit the usual position-space wave-function representation.

To avoid potential subtleties associated with the position-space representation, it is convenient to work in momentum space, where the operator $p$ acts multiplicatively. In this case, we have $q=i\hbar\hspace{0,5mm} \frac{d}{dp}$ and the three representations in Eq. \eqref{threerep} become
\begin{eqnarray}
\nonumber
    &x_1 =\,  i \hbar f(p) \dfrac{\diff}{\diff p},\qquad
	x_2 =\,    i \hbar \left[f(p) \dfrac{\diff}{\diff p} + f'(p)\right]\,,&\\[2mm] 
	&x_3 =\,  i \hbar \left[f(p) \dfrac{\diff}{\diff p} +\dfrac{1}{2} f'(p)\right].&
    \label{sympre}
\end{eqnarray}
where the prime denotes derivative with respect to $p$. 

The representation $x_1$ coincides with the one employed in Ref.~\cite{KMM}. 
Here, however, we adopt the symmetric prescription, corresponding to $x=x_3$; 
for notational simplicity, the subscript ``3'' will be dropped in the following. 
One advantage of this representation is that it allows the standard integration 
measure in momentum space to be retained (see Appendix~\ref{Measure}).


The next step is to implement the reverse quantum/statistical paradigm outlined above. Following the GUP-Tsallis correspondence~\cite{JizbaLuc}, we determine the form of the function $f(p)$ in the quasi-classical regime by requiring consistency between the states saturating Eq.~\eqref{RelatHUP2} and the probability amplitude~\eqref{expk} associated with the $\kappa$-deformed Kaniadakis distribution.

As discussed in~\cite{Jackiw:1968zzb}, the explicit construction of minimum-uncertainty states in the case where the commutator is a $q$-number requires the implementation of an appropriate variational approach. Restricting, without loss of generality, to states centered at the origin of phase space, i.e., satisfying $\langle x\rangle=\langle p\rangle=0$, this procedure implies solving the following differential equation:
\begin{equation}
\label{GenDifEq}
    \left[\frac{x^2}{\Delta x^2} + \frac{p^2}{\Delta p^2} - \frac{2 f({p})}{\langle f({p}) \rangle}\right] |\psi\rangle=0,
\end{equation}
where the state $|\psi\rangle$ saturates the inequality \eqref{RelatHUP2}, namely $\Delta x \hspace{0.2mm}\Delta p=\hbar \langle f(p)\rangle/2$. 

The more general case of displaced states, characterized by non-vanishing expectation values $\langle x\rangle = x_0$ and $\langle p\rangle = p_0$, can be obtained by acting on $|\psi\rangle$ with the unitary operators $\mathcal{S}(p_0)=\exp\!\big(i p_0 q/\hbar\big)$ and $\mathcal{T}(x_0)=\exp\!\big(-i x_0 k\big)$, which generate translations in momentum and position space, respectively. Here, the wave number operator $k$ is defined as the operator canonically conjugate to $x$, satisfying $[x,k]=i$ \cite{BossoLast}. In this case, the differential equation \eqref{GenDifEq} must be consistently modified by the replacements $x^2 \to (x-x_0)^2$ and $p^2 \to (p-p_0)^2$.

A straightforward manipulation permits rewriting Eq.~\eqref{GenDifEq} in the equivalent form
\begin{equation}
    \left(\frac{x}{\Delta x} - i \frac{p}{\Delta p}\right) \left(\frac{x}{\Delta x} + i \frac{p}{\Delta p}\right) |\psi\rangle = 0\,.
    \label{Jack}
\end{equation}
We then impose that the momentum-space probability density associated with $|\psi\rangle$ coincides with the 
$\kappa$-deformed Gaussian in Eq.~\eqref{expk}, namely
\begin{eqnarray}
  \label{psip}
    \psi(p)&=&N\left[\exp_\kappa(-\zeta \hspace{0.2mm}p^2)\right]^{1/2}\\[2mm]
    \nonumber
    &=&N\left(\sqrt{1 + \kappa^{2}\zeta^2 p^{4}} - \kappa\hspace{0.2mm} \zeta\hspace{0.2mm} p^2 \right)^{1/(2\kappa)}\,.
\end{eqnarray}
The normalization is given by $\int_{-\infty}^{\infty}|\psi(p)|^2dp=1$, which implies
\begin{equation}
   N=
\sqrt{
(2+\kappa)\,
\sqrt{\frac{\kappa\zeta}{2\pi}}\,
\frac{\Gamma\!\left(\frac{1}{2\kappa}+\frac{1}{4}\right)}
{\Gamma\!\left(\frac{1}{2\kappa}-\frac{1}{4}\right)}\,,
}
\end{equation}
for any admissible value of $\kappa$, where $\Gamma$ is the Euler Gamma function.

The parameter $\zeta>0$ introduces a characteristic inverse squared momentum scale associated with the $\kappa$–Gaussian CSs. In particular, it controls the intrinsic width of the momentum distribution and thereby labels a continuous family of Robertson-saturating uncertainty states compatible with the $\kappa$-deformed algebra (see Appendix \ref{app}). Further details on its physical interpretation will be provided below.\footnote{
At this stage, $\zeta$ is treated as a free scale parameter. A mathematically equivalent formulation can be obtained by expressing the states \eqref{psip} in terms of the auxiliary momentum $\tilde p$ introduced in Eq.~\eqref{auxquan}, through the rescaling $\tilde p=\sqrt{\zeta}\,p$. In this parametrization, the characteristic momentum scale would be naturally set by the relativistic combination $\kappa\hspace{0.2mm} m\hspace{0.2mm} c$.
\label{fn1}}

We emphasize that, in the present construction, a quadratic dispersion relation is assumed in the $\kappa$-deformed distribution \eqref{psip}, corresponding to the leading-order momentum dependence in the weakly relativistic regime. The characteristic relativistic structure with higher-order corrections is intrinsically contained in the nonlinear form of the $\kappa$-exponential, as follows from the expansion of the square-root term.
Even if the full relativistic dispersion relation were used, its effect would reduce to a renormalization of the coefficients in the perturbative expansion of Eq.~\eqref{psip}, without altering the functional dependence of the result. 

\begin{figure}[t]
\centering\includegraphics[width=0.5\textwidth]{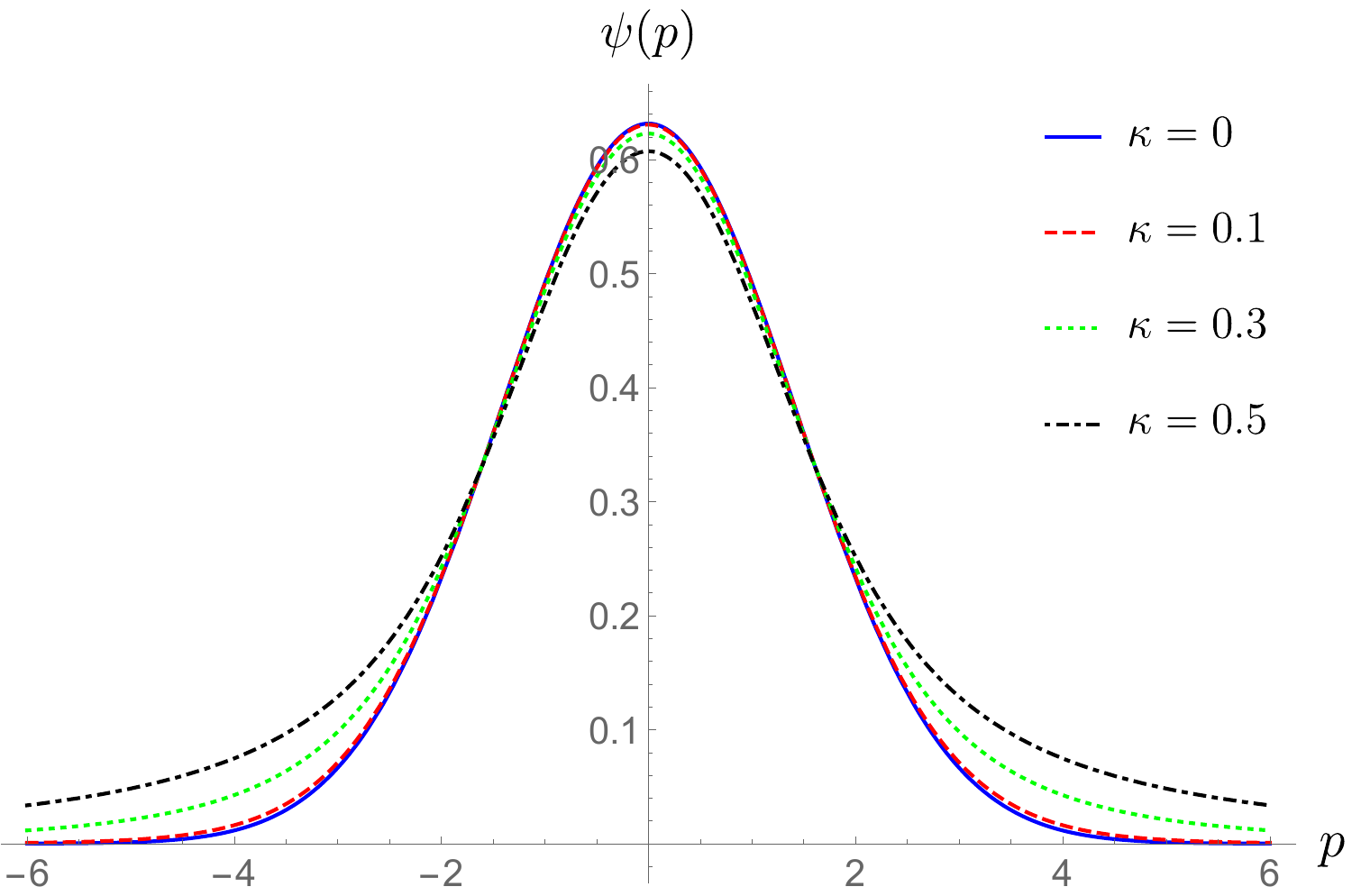}
\caption{Plot of $\psi(p)$ versus $p$, for various values of $\kappa$. We work in natural units.}%
\label{Fig1}%
\end{figure}

The plot of $\psi(p)$ is displayed in Fig. \ref{Fig1} for various values of $\kappa$. It can be seen that the power-law–tailed behavior of the Kaniadakis exponential becomes increasingly pronounced as 
$\kappa$ increases, in contrast to the Gaussian-like decay of the CS that minimizes the HUP ($\kappa=0$, blue solid curve).

By projecting Eq. \eqref{Jack} onto momentum space and making use of Eq. \eqref{sympre}, the resulting differential equation takes the form in Eq. \eqref{difeqps} of Appendix \ref{Measure},  yielding the particular solution 
\begin{equation}
    f(p)=\frac{\Delta x\left(\sqrt{1+\kappa^2\zeta^2 p^4}+\kappa^2\zeta p^2\right)}{\hbar\hspace{0.2mm}\zeta\left(1-\kappa^2\right)\Delta p}+c_1\,\exp_\kappa(\zeta p^2)\,.
    \label{f}
\end{equation}
This expression defines a class of solutions corresponding to different realizations of the algebra compatible with the CS structure \eqref{psip}. 

To select the physically relevant representation, additional consistency conditions must be imposed. In particular, we require that the standard Heisenberg algebra (i.e., $f(p)=1$) be recovered:
\begin{itemize}
    \item[\emph{(i)}] in the limit $\kappa \to 0$, which necessarily implies $c_1 = 0$;
    \item[\emph{(ii)}] in the low-momentum regime ($p \to 0$), which yields 
        $\Delta x = \hbar \zeta \left(1-\kappa^2\right)\Delta p$.
\end{itemize}
These conditions naturally single out the functional form 
\begin{equation}
    \label{fbis}
    f(p)=\sqrt{1+\kappa^2\zeta^2 p^4}+\kappa^2\zeta p^2\,.
\end{equation}
With this choice, the asymptotic behavior of Eq. \eqref{psip},  i.e., $\psi(p)\sim |p|^{-1/\kappa}$ for $|p|\to\infty$,  implies that $\psi(p)$ belongs to the common domain $\mathcal{D}_{x,p}\equiv D(x)\cap D(p)$, provided $0<\kappa<2/3$, where we have defined $D(p):=\{\varphi\in L^2(\mathbb{R}) : p\varphi \in L^2(\mathbb{R})\}$, and similarly for $D(x)$ (see Appendix~\ref{app}). The stronger requirement $\psi\in\mathcal{D}_{x,p,x^2,p^2}$ would instead lead to the more restrictive condition $0<\kappa<2/5$.
Moreover, the operator $x$ is symmetric on a suitable dense domain (e.g., $S_\infty$ \cite{KMM}).
A family of self-adjoint extensions can in principle be obtained by imposing proper boundary conditions, following
a procedure analogous to that outlined in \cite{JizbaLuc}.

Therefore, the saturation condition \eqref{GenDifEq} serves as a structural criterion to fix the deformation function $f(p)$, thereby fully determining the modified commutation relation
\begin{equation}
\label{RUP}
    [x,p]=i\hbar \left(\sqrt{1+\kappa^2\zeta^2 p^4}+\kappa^2\zeta p^2\right).
\end{equation}
This commutator constitutes the main conceptual result of our analysis.
Although reconstructed from the Robertson-saturating sector,
Eq.~\eqref{RUP} is elevated to an operator identity defining the
deformed algebra.
The standard QM commutator is recovered in the limit $\kappa\to0$, corresponding to $c\to\infty$ (see the discussion below Eq.~\eqref{auxquan}).

Let us now focus on the weakly relativistic regime, defined by states whose momentum distribution is concentrated in the domain $\sqrt{\zeta}\,|p|\ll 1$. As discussed in the Introduction, this intermediate domain is precisely the regime relevant to our analysis, since relativistic effects start to become appreciable while still allowing for a controlled perturbative treatment. Furthermore, it provides the most suitable setting for experimentally probing SR corrections to the HUP. In addition, working in this regime allows for a direct comparison with other relativistic extensions proposed in the literature. 

Expanding the operator-valued function $f(p)$ in powers of $\zeta p^{2}$, Eq.~\eqref{RUP} yields
\begin{equation}
\label{apprcom}
    [x,p]=i\hbar\left(1+\kappa^2\zeta p^2\right)+\mathcal{O}\!\left(\kappa^2\zeta^2p^4\right)\,.
\end{equation}
While this relation resembles the quadratic GUP commutator~\cite{KMM}, its physical origin is fundamentally different. In conventional GUP scenarios, the deformation parameter is associated with quantum-gravitational effects and has a geometric interpretation, reflecting the existence of a minimal length scale and the corresponding fuzziness of spacetime at Planckian distances. By contrast, in the present framework the correction arises from the underlying Kaniadakis structure and, in particular, from the non-linear composition laws of relativistic statistical mechanics~\cite{Kania0,Kania1}. Consequently, the two effects are expected to manifest at very different physical scales: the Planck scale in the case of the GUP, versus the relativistic statistical scale associated with the Kaniadakis deformation in the present framework. 

At leading order, and for states symmetric in momentum space, the commutator \eqref{apprcom} implies the uncertainty relation
\begin{equation}
\label{RUPapprox}
    \Delta x\,\Delta p \ge \frac{\hbar}{2}\left(1+\kappa^2\zeta\,\Delta p^2\right),
\end{equation}
which reduces to the standard HUP in the limit $\kappa \to 0$.

\section{Physical Implications}
\label{App}

Let us examine some physical and conceptual implications of the RUP \eqref{RUPapprox}. In particular, we first discuss some of its direct effects and then compare our construction with other relativistic extensions of the HUP.

\subsection{Phenomenological predictions}

\subsubsection{Relativistic minimal length}

As a first observation, we note that, since the RUP \eqref{RUPapprox} has the same structure as the GUP \cite{KMM}, it exhibits formally similar features.
In particular, the quadratic dependence on $\Delta p$ implies a nonvanishing minimal position uncertainty, obtained by minimizing the right-hand side with respect to $\Delta p$. This yields 
\begin{equation}
\label{minunc}
    \Delta x_{\mathrm{min}}=\hbar \kappa \sqrt{\zeta}\,.
\end{equation}
From a Hilbert space perspective, the presence of a minimal position uncertainty modifies the structure of the theory. In particular, exact position eigenstates are no longer physical, as they would require $\Delta x=0$. Instead, one must introduce maximally localized states, which satisfy $\Delta x = \Delta x_{\mathrm{min}}$. These solutions, which replace the usual Dirac position eigenstates as the operational notion of localization, can be constructed following Ref.~\cite{KMM}.

\subsubsection{Constraint from fine-structure constant} 

We now turn to a phenomenological analysis of the RUP \eqref{RUPapprox}, aiming to constrain the Kaniadakis parameter by confronting the modified uncertainty relation with high-precision measurements. In this context, the fine-structure constant provides an ideal testing ground, as its exceptionally precise determination enables stringent bounds on possible deviations from standard quantum theory~\cite{Nasseri1,Lemos}.

To this end, following the approach of Ref.~\cite{Lemos}, it proves convenient to  rewrite Eq. \eqref{RUPapprox} in the Heisenberg-like form $\Delta x \Delta p \ge h_{\mathrm{eff}}/2$, 
where the effective Planck constant is defined by 
\begin{equation}
\label{heff}
h_{\mathrm{eff}}=\hbar \left(1+\kappa^2\zeta\Delta p^2\right).
\end{equation}
Assuming saturation of Eq.~\eqref{RUPapprox}, we solve for the momentum uncertainty and obtain
\begin{equation}
\label{Dpsol}
    \Delta p= \frac{\Delta x-\sqrt{\Delta x^2-\hbar^2\kappa^2\zeta}}{\hbar\kappa^2\zeta}\,,
\end{equation}
where we have retained only the solution that reproduces the standard Heisenberg result in the limit $\kappa\rightarrow0$.

To connect the uncertainty-based estimate \eqref{Dpsol} with atomic physics, one must associate a characteristic spatial scale with the electron in a bound state. In the hydrogen atom, a natural candidate is the Bohr radius $a_0$. Indeed, for the ground state, the radial probability distribution peaks at distances of order $a_0$, and both the expectation value and the root-mean-square radius are proportional to this scale. It is therefore reasonable to assume $\Delta x \sim a_0$\footnote{
Since the analysis of~\cite{Lemos} is formulated within a quantum gravity braneworld framework, the position uncertainty is identified with an effective Bohr radius that incorporates the gravitational corrections induced by the thick-brane geometry.}~\cite{cohen}.

Within this setting, inserting Eq. \eqref{Dpsol} into the definition \eqref{heff} leads to
\begin{equation}
\hbar_{\rm eff}=
\frac{2a_0\left(a_0-\sqrt{a_0^2-\hbar^2\kappa^2\zeta}\right)}
{\hbar\kappa^2\zeta}.
\end{equation}
Therefore, we can introduce an effective fine-structure constant as
\begin{eqnarray}
\nonumber
    \alpha_{\rm eff}=\frac{Q^2}{4\pi\epsilon_0\hspace{0.2mm}\hbar_{\rm eff} c}&=&\alpha \left(\frac{1+\sqrt{1-\hbar^2\kappa^2\zeta/a_0^2}}{2}\right)\\[2mm]
    &\approx&\alpha\left(1-\frac{\hbar^2\kappa^2\zeta}{4a_0^2}\right),
\end{eqnarray}
where $\alpha$ is the standard fine-structure constant. For convenience, we rewrite the above relation as $\alpha_{\rm eff}=\alpha+\Delta \alpha_\kappa$.

To constrain the Kaniadakis-induced correction, we remind that the most precise measurement of the fine-structure constant yields $\alpha^{-1}=137.035999206(11)$~\cite{Measure}, with the experimental uncertainty being expressed in parentheses. Requiring that the predicted shift does not exceed the current experimental sensitivity (i.e., $|\Delta\alpha_\kappa|<\delta\alpha_{\mathrm{exp}}$), we obtain $\kappa\sqrt\zeta \lesssim \mathcal{O}(10^{-3})\,(\mathrm{MeV/c})^{-1}$.

This constraint can be translated into a bound on $\kappa$ by fixing the  scale $1/\sqrt{\zeta}$ to the characteristic electron momentum in the hydrogen ground state, namely $1/\sqrt{\zeta}\simeq3.7\times10^{-3}\,\mathrm{MeV}/c$. This yields the upper bound
\begin{equation}
    \kappa\lesssim\mathcal{O}(10^{-5}).
\end{equation}
It is instructive to place this result in a broader perspective. Existing constraints on $\kappa$ span several orders of magnitude, from nearly vanishing values in black hole thermodynamics and cosmology~\cite{LucianoBH,Rev2,LucianoRev,Hernandez-Almada:2021aiw,She,Drepa,LambiaKan} to values as large as $\kappa \sim 10^{-1}$ inferred in cosmic ray physics~\cite{Kania1}. This hierarchy likely reflects the fact that $\kappa$ acts as an effective parameter whose phenomenological value depends on the  observational context in which it is probed.
In this picture, our bound identifies an intermediate regime where the corrections remain perturbative yet non-negligible, offering a controlled setting to explore extensions of the standard theoretical framework.

\subsection{Comparison with other approaches}
\label{comparison}

In order to further examine the structure of the uncertainty relation \eqref{RUPapprox}, we now compare it with other relativistic generalizations of the HUP discussed in the literature.

\subsubsection{Landau–Peierls approach}
In the seminal analysis of Landau and Peierls~\cite{Land}, an effective modification of the HUP emerges from a consistency requirement: any admissible position measurement must not induce energy fluctuations large enough to trigger particle creation. This condition implies a lower bound on the spatial uncertainty of a particle at rest, of the order of its Compton wavelength. In other words, the impossibility of arbitrarily sharp localization arises from the interplay between the uncertainty principle and the relativistic threshold for pair production. 

In this respect, the present formalism provides an algebraic realization of the same physical limitation identified in \cite{Land}. By identifying the Compton wavelength $\lambda_C$ of a single-particle system with the minimal position uncertainty in Eq.~\eqref{minunc}, the scale $\zeta$ can be fixed as
\begin{equation}
\hbar \kappa \sqrt{\zeta} \sim \frac{\hbar}{mc}
\,\,\Longrightarrow\,\,
\sqrt{\zeta} \sim \frac{1}{\kappa mc}\,.
\label{Landcond}
\end{equation}
Notice that imposing this condition is equivalent to expressing the wave function~\eqref{psip} in terms of the auxiliary momentum $\tilde p$ introduced in Eq.~\eqref{auxquan} (see footnote~\ref{fn1}). In this sense, the Compton-like scaling of $\Delta x_{\mathrm{min}}$ is already implicit in the $\kappa$-deformed parametrization \eqref{psip}. 

Fixing the momentum scale as in Eq.~\eqref{Landcond} also highlights a key conceptual difference between the RUP and the GUP considered in quantum gravity. In the latter case, the minimal position uncertainty is of the order of the Planck length, reflecting the existence of a universal scale below which spacetime itself cannot be operationally probed \cite{KMM}. By contrast, in the present framework the minimal uncertainty is of the order of the Compton wavelength of the system under consideration. The resulting bound is therefore not associated with a universal geometric cutoff, but with the relativistic kinematics of particle localization, and is thus fully compatible with the principles of special relativity.

\subsubsection{Amelino-Camelia and Vestuti approach}

A further interesting comparison can be drawn with the framework of Ref.~\cite{GAC}, where relativistic corrections to the HUP are derived from an operational analysis of the localization procedure. Although both approaches address the intermediate regime in which relativistic effects are small but non-negligible, they differ significantly at the methodological level. 

In Ref.~\cite{GAC}, the relativistic modification arises from an analysis of the measurement process, modeled as the interaction between the system and a probe particle. The deformation is implemented at the level of the space-time probability distribution, without modifying the canonical commutation relations. In particular, the resulting lower bound on $\Delta x$ is consistent with the Landau–Peierls argument \cite{Land}, and, in the regime of relatively low momentum uncertainty, the corresponding relation takes the form
\begin{equation*}
    \Delta x^2 \Delta p^2\gtrsim \frac{\hbar^2}{4}\left(1+\frac{3}{2}\hspace{0.2mm}\frac{\Delta p^2}{m^2c^2}\right).
\end{equation*}

For comparison, squaring our Eq.~\eqref{RUPapprox} and expanding to leading order in $\Delta p^2$, we obtain
\begin{equation}
     \Delta x^2 \Delta p^2\ge \frac{\hbar^2}{4}\Big(1+2\kappa^2\zeta \Delta p^2+\mathcal{O}(\Delta p^4)
     \Big)\,.
\end{equation}
Matching the two expressions at leading order yields $\zeta=\dfrac{3}{4\kappa^2 m^2c^2}$, in agreement with the estimate in Eq.~\eqref{Landcond}, up to a numerical factor of order unity. In this way, our framework promotes the relativistic correction identified in \cite{GAC} from an operational feature of the measurement process to a structural property of the quantum algebra, while remaining consistent with the argument of Ref.~\cite{Land}.

\subsubsection{Putra-Alrizal approach}

Finally, we compare our model with the Relativistic Heisenberg Uncertainty (RHU) proposed in Ref.~\cite{Putra}. In that framework, relativistic effects are incorporated within relativistic QM by analyzing the Lorentz transformation properties of position and momentum operators and exploiting the Ehrenfest theorem to relate expectation values to classical quantities. The resulting bound is explicitly velocity-dependent and reads
\begin{equation*}
\Delta x\,\Delta p \;\ge\; \frac{\hbar}{2}\,\gamma^{2}(v_g),
\qquad
\gamma(v_g)=\frac{1}{\sqrt{1-\frac{v_g^{2}}{c^{2}}}},
\label{Putra_RHU_explicit}
\end{equation*}
where $v_g$ denotes the average (group) velocity of the wave packet. 
In the weakly relativistic regime $v_g\ll c$, this relation can be expanded as
\begin{equation*}
\Delta x\,\Delta p
\;\ge\;
\frac{\hbar}{2}
\left(
1+\frac{v_g^{2}}{c^{2}}
+\mathcal{O}\!\left(\frac{v_g^{4}}{c^{4}}\right)
\right).
\label{Putra_expanded_rigorous}
\end{equation*}
Therefore, the relativistic amplification of the minimal phase-space cell arises directly from the Lorentz factor associated with the transformed observables.

In this case, a term-by-term identification with our RUP \eqref{RUPapprox} is not straightforward without additional assumptions relating the group velocity $v_g$ to the momentum distribution of the wave packet. Thus, although both constructions predict a relativistic enhancement of the Heisenberg bound, they differ in their conceptual and structural foundations: in Ref.~\cite{Putra} the effect is governed by the mean kinematical quantity $v_g$, whereas in the present model it is controlled by the intrinsic quantum spread $\Delta p$ through a deformed commutation structure rooted in Kaniadakis statistics.

\section{Discussion and Conclusions}
\label{D&C}
In this work, we have proposed a relativistic extension of the HUP grounded in the framework of Kaniadakis statistics. Building on a quantum–statistical paradigm previously established for Tsallis entropy \cite{JizbaLuc}, we reconstructed the deformation of the canonical commutator by enforcing consistency between the structure of the minimum uncertainty states and the probability amplitude of the Kaniadakis distribution. The resulting algebra, Eq.~\eqref{RUP}, provides an exact realization of the Relativistic Uncertainty Principle (RUP), promoting the relativistic correction to a modification of the  commutator. At a conceptual level, our result is reminiscent of $k$-deformed relativistic frameworks, in which relativistic effects are associated with deformations of the underlying algebraic structures. Although the present model does not implement a Hopf-algebra deformation of relativistic symmetries, it can nevertheless be interpreted as an effective relativistic deformation of the Heisenberg algebra~\cite{Lukierski:1991pn,Majid:1994cy}.

In the perturbative regime, our commutator yields a quadratic correction in $\Delta p$ to the standard uncertainty relation, formally analogous to that arising in GUP models. However, this correspondence holds only at leading order. The full structure is encoded in the exact $\kappa$-deformed algebra~\eqref{RUP}, whose origin and conceptual basis differ fundamentally from quantum-gravity-motivated GUP frameworks.

We further explored physical implications of the RUP. In particular, we derived phenomenological constraints on the Kaniadakis parameter from precision measurements of the fine-structure constant, obtaining the bound $\kappa \lesssim \mathcal{O}(10^{-5})$. We also compared our construction with other relativistic extensions of the HUP, highlighting both common features in the weakly relativistic regime and essential differences in their underlying structure.

Several directions remain open for future investigation. From a theoretical standpoint, a fully relativistic generalization of the uncertainty principle would require a manifestly Lorentz-invariant formulation in which space and time are treated on equal footing, possibly through a four-vector realization of the commutation relations or a deformation of relativistic phase space. 

Another interesting direction would be to explore whether the present construction can be embedded within more general entropic frameworks that interpolate between different extensions of the Boltzmann–Gibbs entropy \cite{Nojiri:2022dkr}, potentially providing a unified statistical setting in which different deformations of the uncertainty principle emerge as limiting regimes.

From an experimental perspective, it would be important to assess whether the leading-order correction derived here can be tested in high-precision quantum experiments operating in regimes where relativistic effects are small but non-negligible. Further developments along these lines are currently under investigation and will be reported elsewhere.

\appendix

\section{Measure in momentum space}
\label{Measure}

Let us note that the three representations in Eq.~\eqref{sympre} can be expressed in the compact form
\begin{equation}
\label{A1eq}
x = i \hbar \left[f(p) \frac{\diff}{\diff p} + A f'(p)\right] 
  = i \hbar [f(p)]^{1-A} \frac{\diff}{\diff p} [f(p)]^{A}.
\end{equation}
The three cases in Eq.~\eqref{sympre} correspond, respectively, to 
\begin{equation}
A=0\,\, (x_1), \quad\, A=1\,\, (x_2), \quad\, A=\frac{1}{2}\,\, (x_3)\,.
\end{equation}
In the latter case, which is the choice adopted in the present analysis, the differential equation \eqref{Jack}, with $\psi(p)$ given in Eq.~\eqref{psip}, takes the form
\begin{eqnarray}
\label{difeqps}
&&\hspace{-4mm}4\hbar^2 \zeta\Bigl[1-p^2\zeta(\kappa^2 p^2\zeta+s(p))\Bigr]
\Delta p^2\, f(p)^2 \\[2mm]
&&\hspace{-2mm}+\, s(p)^3\Bigl(4p^2\Delta x^2-\hbar^2\Delta p^2\,f'(p)^2\Bigr)
+2\hbar s(p)^2\Delta p\,f(p) \nonumber\\[2mm]
\nonumber
&&\hspace{-2mm}
\times\,\Bigl[
4\hbar p\zeta\,\Delta p\,f'(p)
-s(p)\Bigl(2\Delta x+\hbar\Delta p\,f''(p)\Bigr)
\Bigr]=0\,.
\end{eqnarray}
where we have introduced the shorthand notation
\begin{equation}
s(p)\equiv\sqrt{1+\kappa^2\zeta^2 p^4}\,.
\end{equation}

Now, following the reasoning of Ref.~\cite{KMM}, we determine the appropriate
measure in momentum space that renders the operator ${x}$ symmetric. In fact, we observe that, with the standard momentum-space measure, neither $x_1$ nor $x_2$ in are symmetric, whereas $x_3$ is. 
To show this, we introduce a real weight function $g(p)$ defining a
modified integration measure and consider the action of the operators on the
dense domain $S_\infty$ of functions that decay faster than any power
of $p$~\cite{KMM}. 

Under these assumptions, we obtain
\begin{eqnarray}
   && \int_{-\infty}^{\infty}\diff p ~ g(p) \phi_1^\star (p) \,x\, \phi_2(p)\\[2mm]
    \nonumber
    &&\hspace{2mm}= i \hbar \int_{-\infty}^{\infty}\diff p ~ g \phi_1^\star \left[f \frac{\diff}{\diff p} + A f'\right] \phi_2\\[2mm]
&&\hspace{2mm}= 
    \int_{-\infty}^{\infty} \diff p ~ \left\{ - i \hbar A g f' \phi_1 + i \hbar \frac{\diff}{\diff p} \left[g f \phi_1\right]\right\}^\star \phi_2, 
\end{eqnarray}
where we performed a partial integration. 
Then, we require that
\begin{equation}
- i \hbar A g f' \phi_1 + i \hbar \frac{\diff}{\diff p} \left[g f \phi_1\right]= g\, i \hbar \left[f \frac{\diff}{\diff p} + A f'\right] \phi_1.
\end{equation}
Imposing the additional condition that standard QM
is recovered in the low-momentum limit, \emph{i.e.} $g(0)=1$, we obtain
\begin{equation}
    g(p) = [f(p)]^{2A-1}.
\end{equation}
For $A=0$ one recovers the measure employed in Refs.~\cite{KMM}. Furthermore, for $A=\frac{1}{2}$
one has $g(p)=1$ for all $p$. This is consistent with the fact that
$x_3$ is symmetric with respect to the standard integration
measure in momentum space.

In terms of the momentum eigenstates $|p\rangle$, the identity operator
can therefore be written as
\begin{equation}
\label{ident}
\mathds{1}=\int_{-\infty}^{\infty}dp~[f(p)]^{2A-1}|p\rangle\langle p|\,,
\end{equation}
which, for $A=0$, reduces to the completeness relation adopted in
Ref.~\cite{KMM}.

As a final point, following the discussion in \cite{BossoLuciano}, it is useful to note that the wavefunctions
corresponding to different operator orderings, labeled by the
parameter $A$, are not independent but can be related through a
simple transformation. To illustrate this, let us consider the
stationary Schr\"odinger equation in momentum space for a generic
potential $U(x)$,
\begin{equation}
    \frac{p^2}{2m}\,\phi^{(A)}(p)
    + \bigl(U(x)-E\bigr)\phi^{(A)}(p)=0\,.
    \label{eqn:schrodinger}
\end{equation}

Using the representation of the position operator introduced above,
one finds that repeated action of $x$ on the wavefunction can be
expressed as
\begin{equation}
x^n \phi^{(A)}(p)
=
\frac{(i\hbar)^n}{[f(p)]^{A}}
\left(f(p)\frac{\diff}{\diff p}\right)^n
\!\left([f(p)]^{A}\phi^{(A)}(p)\right).
\end{equation}
If the potential can be expanded in powers of $x$, it is
convenient to introduce the auxiliary function
\begin{equation}
\label{trans}
\phi^{(0)}(p)=[f(p)]^{A}\phi^{(A)}(p)\,,
\end{equation}
which corresponds to the ordering $A=0$.

In terms of $\phi^{(0)}(p)$ the Schr\"odinger equation takes the form
\begin{equation}
    \frac{p^2}{2m}\,\phi^{(0)}(p)
    +\bigl(U(x)-E\bigr)\phi^{(0)}(p)=0\,.
\end{equation}
Once this equation has been
solved, the wavefunction associated with any other ordering can be
obtained by inverting the transformation \eqref{trans}. In this way, the
solutions corresponding to different operator orderings are related
through a simple rescaling.

This observation also clarifies the structure of the corresponding
Hilbert space. As discussed above, for $A=\tfrac12$ the measure in
momentum space reduces to the standard one, and the associated
Hilbert space coincides with the usual space $L^2(a,b)$ of
square-integrable functions on a (possibly infinite) interval
$(a,b)$. 

From the definition of $\phi^{(0)}(p)$ it follows that
\begin{eqnarray}
\nonumber
&&\phi^{(0)}(p)=[f(p)]^{1/2}\phi^{(1/2)}(p)\\[2mm]
&&\hspace{3mm}\Rightarrow
\phi^{(A)}(p)=[f(p)]^{1/2-A}\phi^{(1/2)}(p).
\label{eqn:wf_conversion_order}
\end{eqnarray}
Therefore, for a given ordering parameter $A$, the corresponding
physical states $\phi^{(A)}(p)$ span the Hilbert space
\[
\mathcal H
=
\left\{
\phi^{(A)}(p):
[f(p)]^{A-1/2}\phi^{(A)}(p)\in L^2
\right\}.
\]

\section{Minimum-uncertainty $\kappa$-Gaussian states}
\label{app}

To assess the self-consistency of the formalism in Sec.~\ref{RHUP}, we examine the structure of the uncertainty relation implied by the deformed algebra \eqref{RUP}. In particular, we verify that the $\kappa$–Gaussian states in Eq.~\eqref{psip} saturate the  Robertson inequality $\Delta x \Delta p \ge \frac{\hbar}{2}\langle f(p)\rangle$.

First, as follows from Eq.~\eqref{GenDifEq}, one can check that these states are centered at the origin of phase space and satisfy $\langle x\rangle=\langle p\rangle=0$. 
As for the expectation value of $p^2$, we get
\begin{eqnarray}
    &&\langle p^2\rangle=\int_{-\infty}^{\infty}p^2|\psi(p)|^2\\[2mm]
    \nonumber
&&\hspace{3mm}=\frac{2+\kappa}{4\kappa\hspace{0.2mm}\zeta\,(2+3\kappa)}
\frac{
\Gamma\!\left(\frac{1}{2\kappa}-\frac{3}{4}\right)
\Gamma\!\left(\frac{1}{2\kappa}+\frac{1}{4}\right)
}{
\Gamma\!\left(\frac{1}{2\kappa}+\frac{3}{4}\right)
\Gamma\!\left(\frac{1}{2\kappa}-\frac{1}{4}\right)
},\quad 0<\kappa<2/3\,,
\end{eqnarray}
which implies for the momentum uncertainty
\begin{equation}
\label{Dp}
    \Delta p =\sqrt{\langle p^2\rangle}=
    \sqrt{\frac{2+\kappa}{4\kappa\hspace{0.2mm}\zeta\,(2+3\kappa)}
\frac{
\Gamma\!\left(\frac{1}{2\kappa}-\frac{3}{4}\right)
\Gamma\!\left(\frac{1}{2\kappa}+\frac{1}{4}\right)
}{
\Gamma\!\left(\frac{1}{2\kappa}+\frac{3}{4}\right)
\Gamma\!\left(\frac{1}{2\kappa}-\frac{1}{4}\right)
} }.
\end{equation}
In turn, using Eqs.~\eqref{sympre} and \eqref{fbis}, we obtain
\begin{eqnarray}
\hspace{-1mm}    \langle x^2\rangle&\hspace{-1.2mm}=\hspace{-1.2mm}&\hbar^2\zeta^2\left(1-\kappa^2\right)^2\langle p^2\rangle\,,\\[2mm]
    \label{Dx}
   \Delta x &\hspace{-1.2mm}=\hspace{-1.2mm}&\sqrt{\langle x^2\rangle}\\[2mm]
    \nonumber
&\hspace{-1.2mm}=\hspace{-1.2mm}&\hbar\sqrt{\zeta}\left(1-\kappa^2\right)
\sqrt{
\frac{2+\kappa}{4\kappa\left(2+3\kappa\right)}
\,
\frac{
\Gamma\!\left(\frac{1}{2\kappa}-\frac{3}{4}\right)
\Gamma\!\left(\frac{1}{2\kappa}+\frac{1}{4}\right)
}{
\Gamma\!\left(\frac{1}{2\kappa}+\frac{3}{4}\right)
\Gamma\!\left(\frac{1}{2\kappa}-\frac{1}{4}\right)
}
}.  
\end{eqnarray}
Equations~\eqref{Dp} and \eqref{Dx} are fully consistent with condition $\emph{(ii)}$ below Eq.~\eqref{f}. Furthermore, the uncertainty product takes the form
\begin{equation}
\label{sateq}
\Delta x\,\Delta p=\frac{\hbar}{2}\,\mathcal{F}(\kappa),
\end{equation}
where we have defined
\begin{equation}
\label{FK}
    \mathcal{F}(\kappa)\equiv
\frac{(1-\kappa^2)}{2\kappa}\,
\frac{\Gamma\!\left(\frac{1}{2\kappa}-\frac{3}{4}\right)\,
\Gamma\!\left(\frac{1}{2\kappa}+\frac{5}{4}\right)}
{\Gamma\!\left(\frac{1}{2\kappa}+\frac{7}{4}\right)\,
\Gamma\!\left(\frac{1}{2\kappa}-\frac{1}{4}\right)}.
\end{equation}
The equality~\eqref{sateq} provides the desired saturation condition. Indeed, using the definition~\eqref{fbis} together with the identity $\Gamma(z+1)=z\Gamma(z)$, one finds that the expectation value of $f(p)$ evaluated on the state $\psi(p)$ in Eq.~\eqref{psip} is precisely $\langle f(p)\rangle=\mathcal{F}(\kappa)$. 

Therefore, for any normalized state $\phi\in\mathcal D$, where $\mathcal D$ is a common dense domain of $x$ and $p$ on which the commutator \eqref{RUP} is well defined, one has \cite{Jackiw:1968zzb}
\begin{equation}
\frac{\Delta x_\phi\,\Delta p_\phi}{\hbar\,\langle f(p)\rangle_\phi}
\;\ge\;
\min_{\varphi\in \mathcal D}
\frac{\Delta x_\varphi\,\Delta p_\varphi}{\hbar\,\langle f(p)\rangle_\varphi}
\;=\;
\frac{\Delta x_\psi\,\Delta p_\psi}{\hbar\,F(\kappa)}=\frac{1}{2}\,,
\end{equation}
consistently with the choice of the algebra \eqref{fbis}.


Furthermore, by considering the asymptotic behavior of the Gamma functions in Eq. \eqref{FK}, one readily finds that $\lim_{\kappa\rightarrow 0} \mathcal{F}(\kappa)=1$, thereby recovering the saturation of the HUP in the undeformed limit.

Finally, it is important to emphasize that the constancy of the uncertainty product \eqref{sateq} for the present class of states follows directly from the specific functional form of the wavefunction~\eqref{psip}. In particular, within the CS family considered here, the momentum uncertainty $\Delta p$ is uniquely determined by $\zeta$ and $\kappa$, so that the combination $\zeta\,\Delta p^2$ becomes a function of $\kappa$ only (see Eq.~\eqref{Dp}). Substituting this relation into condition $(\emph{ii})$ below Eq.~\eqref{f}, one finds that the product $\Delta x\,\Delta p$ depends exclusively on $\kappa$, and is therefore constant within this family of states.

This behavior closely parallels that of standard Gaussian CSs in ordinary QM, where the uncertainty product remains fixed at $\hbar/2$ independently of the width of the wave packet. In the present case, the Kaniadakis deformation modifies the numerical value of the minimum uncertainty product through the function $F(\kappa)$, while preserving its state-independent character within the class of relativistic CSs defined by Eq.~\eqref{psip}.

\begin{acknowledgments}
We thank Giorgio Kaniadakis for insightful discussions and helpful comments on the manuscript. The research of GGL is supported by the postdoctoral fellowship program of the University of Lleida. GGL also acknowledges the LISA Cosmology Working Group (CosWG), as
well as support from the COST Actions CA21136 - \textit{Addressing
observational tensions in cosmology with systematics and
fundamental physics (CosmoVerse)} - CA23130, \textit{Bridging high
and low energies in search of quantum gravity (BridgeQG)} and
CA21106 - \textit{COSMIC WISPers in the Dark Universe: Theory,
astrophysics and experiments (CosmicWISPers)}.
\end{acknowledgments}

\end{document}